# Analysis of complex circadian time series data using wavelets

Christoph Schmal[1]*, Gregor Mönke[2], Adrián E. Granada[3,4]

[1] Institute for Theoretical Biology, Humboldt Universität zu Berlin, Philippstr. 13, 10115 Berlin, Germany

[2] European Molecular Biology Laboratory, Meyerhofstraße 1, 69117 Heidelberg, Germany

[3] Charité Comprehensive Cancer Center, Charité Universitätsmedizin Berlin, Charitéplatz 1, 10117 Berlin, Germany.

[4] German Cancer Consortium (DKTK), German Cancer Research Center (DKFZ), 69120 Heidelberg, Germany

*Corresponding author: christoph.schmal@hu-berlin.de

## Summary

Experiments that compare rhythmic properties across different genetic alterations and entrainment conditions underlie some of the most important breakthroughs in circadian biology. A robust estimation of the rhythmic properties of the circadian signals goes hand in hand with these discoveries. Widely applied traditional signal analysis methods such as fitting cosine functions or Fourier transformations rely on the assumption that oscillation periods do not change over time. However, novel high-resolution recording techniques have shown that, most commonly, circadian signals exhibit time-dependent changes of periods and amplitudes which cannot be captured with the traditional approaches. In this chapter we introduce a method to determine time-dependent properties of oscillatory signals, using the novel open-source Python-based *Biological Oscillations Analysis Toolkit* (pyBOAT). We show with examples how to detect rhythms, compute and interpret high-resolution time-dependent spectral results, analyze the main oscillatory component and to subsequently determine these main components time-dependent instantaneous period, amplitude and phase. We introduce step-by-step how such an analysis can be done by means of the easy-to-use point-and-click graphical user interface (GUI) provided by pyBOAT or executed within a Python programming environment. Concepts are explained using simulated signals as well as experimentally obtained time series.

# 1. Introduction

Circadian oscillations are present at all scales of an organism, from the cellular up to the behavioral level. Recent improvements in the experimental techniques have allowed unprecedented long-term high-resolution recordings in cultures of individual cells, *ex-vivo* tissues and even *in-vivo* from freely moving animals [1]–[4]. In some cases, these rhythms show robust stable oscillations with steady period and amplitude, but most frequently they show time-dependent fluctuations in period, amplitude and trends. Nevertheless, when it comes to quantifying these rhythms, the vast majority of the circadian community still relies on software tools that, at their core, rely on methods designed under the premise that oscillations have static time-independent components, also known as the stationarity assumption. Among others, stationarity-based methods implemented ubiquitously in analytical software tools include the well-known Fourier transformations, Lomb Scargle periodograms and cosinor analysis [5].

For specific cases, the biological data analysis community has developed data-analysis tools tailored to the characterization of nonstationary oscillatory components [6], [7]. These new set of robust software solutions are in practice incorporated as an additional step within larger data analysis pipelines that typically include pre-processing denoising, detrending and normalization. We have recently shown that such multi-step pipelines that combine pre-processing steps with a subsequent analysis of oscillatory properties can lead to significant spectral artifacts that often remain undetected [8]. In this chapter, we describe through examples how to use a recently published multi-step open-source software tool, pyBOAT, that integrates all required steps for the analysis of raw circadian data. PyBOAT implements wavelet analysis and was specifically designed for noisy non-stationary datasets that by-design overcomes potential spectral artifacts of the most frequent pre-processing steps in time series data analysis.

In Section 2 of this chapter we provide a set of online sources to download and install pyBOAT. In Section 3.1, we describe the graphical user interface to carry out a spectral analysis and generate figures of the results. Finally, in Section 3.2, we introduce a flexible scripting-based implementation of pyBOAT.

## 2. Materials

Software:

1. pyBOAT is a freely available open-access software that runs on multiple mainstream operating systems such as Linux, MacOS and Windows. It requires a Python 3.x version to be installed on the system.

2. A convenient approach to install Python together with pyBOAT is by means of Anaconda, an open-source Python and R programming language distribution that aims at simplifying package management for scientific computing. Anaconda has a graphical user interface (GUI), the Anaconda Navigator, and thus requires no use of the command-line. An installation manual for Anaconda can be found on: https://docs.anaconda.com/anaconda/install/

3. A detailed guide how to install pyBOAT using Anaconda can be found at

    https://github.com/tensionhead/pyBOAT

    or by following the steps in the pyBOAT's video installation tutorial

    http://granadalab.org/media/

4. In case Anaconda or its package manager conda is already installed on the machine, pyBOAT can be installed via the command line by typing
    ```
    conda config --add channels conda-forge
    conda install pyboat
    ```

5. pyBOAT can be installed without conda using the package-management system pip by typing

   `pip install pyboat`

   into the command-line.

## 3. Methods

### 3.1 Graphical User Interface

pyBOAT contains an easy-to-use graphical user interface (GUI) that requires no programming experience. In the following paragraphs we will introduce step-by-step how to perform a wavelet analysis of circadian time series data using the GUI of pyBOAT. In order to illustrate the strength of our wavelet analysis approach, we will first investigate a simulated oscillatory time series with well-known properties. Experimental data is then subsequently analyzed in Sections 3.1.8 and 3.2.6.

#### 3.1.1 Example data

Commonly applied time series analysis methods such as Fourier analysis, Lomb-Scargle periodograms or fitting of harmonic functions reliably estimate the period of main oscillatory components as long as oscillatory properties remain stable or vary little over time. In contrast, time-frequency methods such as wavelet analysis are well suited to uncover time dependent (i.e. instantaneous) periods and amplitudes. In order to illustrate the strength of wavelet analysis, we analyze a simulated non-stationary, noisy oscillatory time series whose period changes linearly from 22h to 26h within a week. This synthetic signal additionally exhibits a non-linear baseline expression trend as well as a decaying amplitude, signal properties often encountered in practice.

#### 3.1.2 Download example data

In Section 3.2.1 we describe how to simulate and save the example time series data that we are going to analyze in the next paragraphs, using Python commands. Alternatively, the data can be

downloaded from the following link:

https://github.com/cschmal/chapter-wavelets

### 3.1.3 Launch pyBOAT

The graphical user interface of pyBOAT can be started either by launching it from the Anaconda Navigator or by typing `pyboat` into the command line.

### 3.1.4 Import data

The main window of pyBOAT (Figure 1A) contains three elements. The *Start Generator* button on the right column launches a signal generator to analyze synthetic (simulated) signals (*see* **Note 1**). This can be useful for teaching purposes or to accommodate with the properties and features of pyBOAT and its underlying wavelet analysis. The buttons within the left column allow to import external data:

1. pyBOAT expects tabular data in one of the supported formats `.xls, .xlsx, .csv, .tsv` or `.txt`. Each column should contain a time series signal, sampled at equidistant time points (*see* **Note 2**).

2. Click on the *Open* button and select the file that contains the data to be imported. Column names are automatically inferred from the first row of the data file. We can use this option to import the data set from Section 3.1.2.

3. Click on the *Import* button for more importing options as shown in Figure 1B. For non-standard data formats, uncheck *Separator from extension* and specify a custom *Column separator* in the corresponding box. In case the first row of the data file does not contain column names, check the *No header row present* box. In addition, you can choose to check the *Interpolate missing values* option for a linear interpolation of gaps in your data set (*see* **Note 3**).

4. After importing the data, the *DataViewer* window opens as shown in Figure 1C. Within the *DataViewer* you can see the head of the imported data and set various options for the subsequent analysis.

5. First, choose a data column for further analysis. We will analyze the second column since the first column contains the time points of data sampling in our data set. This column can be chosen either by clicking on the respective column or by choosing the column name within the *Select Signal* box.

6. Second, we need to specify the sampling interval and corresponding time unit. In our example we analyze (simulated) data that has been sampled at a 15min interval. Thus, we choose "0.25" within the *Sampling Interval* box and "hours" within the *Time Unit* box, see upper part of Figure 1C.

7. Click the *Refresh Plot* button. One should now see the raw time series signal for the chosen time axis units in the *Signal and Trend* plotting window (Figure 1C *bottom left*).

**3.1.5 Detrending**

Circadian time series often exhibit long term changes in their magnitude of oscillation. While changes in this magnitude can be informative by themselves, it is often useful to remove this baseline trend for a better representation of oscillatory components and further analysis. In *Mönke et al.* [8] we argue that the sinc filter is a good choice for removing nonlinear trends (low frequency components) of oscillating time series, while minimizing common detrending artifacts such as spurious oscillations. The sinc filter works as a step function in the frequency domain. It removes signal components that are larger than a predefined cut-off period and neither attenuates nor amplifies components below this cut-off period [8].

1. Choose a cut-off period for the sinc filter by typing a numerical value into the *Cut-off Period* box of the *Sinc Detrending* panel within the *DataViewer*. Here, we chose a cut-off period of

48h, i.e. roughly twice as long as the expected period of the circadian signal (Figure 1C, *see* **Note 4**).

2. Check the box *Trend* within the *Plotting Options* of the *DataViewer* to plot the trend, determined by the sinc filter. The nonlinear parabola shaped trend of this synthetic time series is nicely captured (Figure 1C *purple line*).

3. One can plot the detrended time series by additionally checking the *Detrended Signal* box.

4. The raw data, trend and detrended time series data can be saved into a three-column data file via the *Save Filter Results* button. Supported output formats are *.txt, *.csv or *.xlsx.

### 3.1.6 Analysis and detection of periodic signals using wavelets

1. For wavelet-based time frequency analysis, a mother wavelet (*see* **Note 5**) probes the signal of interest along the time axis for a range of pre-defined frequencies or periods. This range of periods of interest has to be supplied by the user within the *Analysis* panel of the *DataViewer*. One can specify the periods to be analyzed between a *Lowest period* and a *Highest period* at equidistant steps for a given total *Number of periods* by typing numerical values into the corresponding boxes. If there are periods present in the signal outside of this pre-defined range, pyBOAT cannot detect them.

2. Here, we chose a lowest period of 0.5h, tantamount to the Nyquist period given by two-times the sampling interval. The Nyquist period is also the default value used by pyBOAT (see Figure 1C).

3. In order to perform the wavelet analysis on the sinc-detrended time series one has to check the *Use Detrended Signal* box.

4. Click on the *Analyze Signal* button to perform the wavelet analysis.

5. After the computation is done, the *Wavelet Spectrum* window opens, see Figure 2A. The upper part of the window shows the analyzed signal, i.e. in our case the detrended time series. The middle part of the window shows the wavelet spectrogram, the main result of our time frequency analysis. Such spectrogram gives a detailed time resolved picture that is able to unveil time-dependent oscillatory properties (see Figure 2A) as well as multiple oscillatory components such as ultradian rhythms subordinated to circadian oscillations [7].

6. The region of maximum power shows a clear trend towards longer periods for later times *t* and thus successfully captures the linear evolution from a period of 22h up to 26h within one week as described in Section 3.1.1. The decaying power for later times t reflects the decaying amplitude of the signal.

### 3.1.7 Ridge analysis reveals the main rhythmic component

Although the wavelet spectrogram gives a complete picture of the time-resolved oscillatory properties, potentially including multiple dominant periods, one is often interested in identifying a main oscillatory component and its properties. Such main oscillatory component can be deduced from a wavelet ridge (*see* **Note 6**). pyBOAT connects the set of maximal power values in the spectrogram along successive time points to determine the ridge:

1. Click on the *Detect Maximum Ridge* button within the *Ridge Detection* panel of the *Wavelet Spectrum* window to compute the maximum power ridge. Subsequently, a bold red line depicts the ridge within the wavelet spectrogram (see Figure 2A).

2. In order to avoid evaluation of the ridge within a low spectral power (noise) regime one can set a minimum wavelet power threshold (*see* **Note 7**) by typing a numerical value into the *Ridge Threshold* box of the *Ridge Detection* panel.

3. To work around sudden jumps within the ridge, e.g. due to poor spectral resolution of the transform, the ridge can also be smoothed by choosing a Savitzky-Golay window size.

4. Click on the *Plot Ridge Readout* button to evaluate the time-dependent (instantaneous) period, amplitude, power and phase of the main oscillatory component. A new window termed *Wavelet Results* will subsequently open, see Figure 2B.

5. The *Wavelet Results* window shows the time-dependent oscillation period (*upper left*), phase (*upper right*) and amplitude (*bottom left*) as well as the spectrogram power along the maximum ridge (*bottom right*). Values within or outside the cone of influence are depicted by *dashed* or *bold lines*, respectively (*see* **Note 8**).

6. To save these results click on *Save Results* at the bottom of the window. Supported formats are `*.txt`, `*.csv` and `*.xlsx`. See Figure 2C for an example readout.

**3.1.8 Ensemble analysis**

The GUI of pyBOAT provides a convenient way to analyze large ensembles of time series data (*see* **Note 9**) and gives various summary statistics such as the ensembles period and amplitude distribution or phase coherence. We demonstrate this functionality using a PER2::LUC bioluminescence recording within coronal slices of the mammalian central pacemaker - the suprachiasmatic nucleus (SCN) - as previously published in *Abel et al.* [1]. Within this data set, SCN slices have been treated with tetrodotoxin (TTX) to suspend spike-associated couplings, four days after starting the *in vitro* recordings. After another six days, TTX has been washed out from the medium. Circadian oscillatory signals of individual SCN neurons could be identified and tracked throughout the time lapse recordings, see *Abel et al.* for further details. Within the next steps we study the effect of TTX on dynamical properties of the SCN neurons circadian PER2::LUC oscillations using the batch analysis function of pyBOAT:

1. PER2::LUC oscillatory time series and the corresponding locations of SCN neurons within the time lapse recordings of *ex vivo* SCN slices as published in *Abel et al.* can be downloaded from: https://github.com/JohnAbel/scn-resynchronization-data-2016

2. Here, we chose the data set 'scn2_full_data.csv'. It contains 264 times series recorded from the tracked individual SCN neurons. Since the file lacks descriptive headers in the first row, we import the data via the *Import* button of pyBOATs main screen and check the *No header row present* box.

3. After importing the data, we set the *Sampling Interval* to 1 and the *Time Unit* as 'h' as used in the experimental protocol, see Figure 3A.

4. Amplitudes of PER2::LUC oscillations drastically decrease upon TTX treatment. To reduce edge effects at this transition for sinc filter detrending we choose a relatively low filter cut-off of 36h (*see* **Note 10**), compare Figure 3A.

5. Once suitable parameters for sinc detrending and the subsequent wavelet analysis have been found (compare Figure 3A-B) one can run the batch or ensemble analysis by clicking on the *Analyze All* button.

6. Choose the *Summary Statistics* of interest, *Ridge Detection* parameters as described above and required export options within the *Batch Processing* window.

7. Check the *Ensemble Dynamics* box in the *Summary Statistics* panel and click on the *Run for 264 Signals* button to execute the batch analysis.

8. The *Ensemble Dynamics* window depicts the mean and standard deviation of the time-dependent (instantaneous) period, amplitude and wavelet spectrum power as well as the phase coherence for the whole ensemble of signals. It can be seen that application of TTX reversibly broadens the period distribution (i.e. standard deviation increases), decreases the phase coherence and leads to a decrease in amplitude which can be indicative for a reduced relative coupling strength as described previously [9]–[11].

**3.2. Implementing pyBOAT within a Python Script**

Even though pyBOAT has an easy-to-use graphical user interface (GUI), it can be more convenient in some cases to run the analysis routines of pyBOAT within a Python script. This applies for example to cases where the wavelet analysis provided by pyBOAT is only part of a larger analysis pipeline or to cases of extremely long time series data where, due to the long computation time of wavelet spectra, the practitioner could want to run the analysis on an external computing cluster.

In the following sections we demonstrate how to analyze time series data with pyBOAT using the Python programming language. The code can be run either by executing each line in an interactive command shell such as IPython or a Jupyter Notebook, or by copying the code into a file - e.g. "example.py" - and running it in the shell using the command `python example.py`

### 3.2.1 Generating simluated complex oscillatory time series data

The following steps show how to create and save the example time series data as analyzed in Sections 3.1.4 – 3.1.7.

1. The NumPy scientific computing library for Python provides convenient ways to simulate rhythmic time series. The library can be imported via

   ```
   import numpy as np
   ```

2. We define an equidistant set of time points, using a sampling interval of 15min for an overall length of one week:

   ```
   dt   = 0.25                           # Sampling interval in hours
   tvec = np.arange(0, 7 * 24 + dt, dt)   # Time array
   ```

   Please note that the Python interpreter ignores everything that follows a "#" within a given line.

3. We next define a time-dependent (instantaneous) period of the simulated time series that linearly lengthens from 22h to 26h within one week:

   ```
   T = 22. + 4. / (7. * 24.) * tvec
   ```

4. An oscillatory signal based on this instantaneous period T is constructed via the NumPy cosine function:

```
signal  = np.cos(2 * np.pi / T * tvec)
```

5. In many cases, such as bioluminescence recordings in cell culture or tissue slices, the corresponding circadian time series exhibit a decay in amplitude. We model this by introducing an exponential decay at a half-life of 48h:

```
signal  = signal * np.e**(-np.log(2) / 48. * tvec)
```

6. In addition, many experimental circadian time series show a non-linear trend in baseline expression or magnitude. Here, we add a simple non-linear baseline trend to our signal, given by a mirrored parabola

```
signal  += -tvec * (tvec – 8 * 24) * 0.0002
```

7. Noise is omnipresent in biological signals and can be due to uncertainties in the measurement process or due to the intrinsic probabilisitic nature of biological processes. Noise can confound otherwise precise deterministic observables of interest but can, on the other hand, also be a critical function to drive biological processes [12], [13]. Here, we add for illustrative purposes uncorrelated (Gaussian) white noise to our simulated time series by adding an array of normally distributed random numbers of standard deviation 0.15 and of mean 0, using the "normal" function of NumPy's "random" package:

```
signal  += np.random.normal(loc=0, scale=0.15, size=len(tvec))
```

The impact of different kinds of noise as well as noise strength are discussed in more detail in *Mönke et al.* [8].

### 3.2.2 Import and initialization of the wavelet analyzer

1. Import the wavelet analyzer of pyBOAT by typing

    ```
    from pyboat import WAnalyzer
    ```

2. Analogously to Section 3.1.6. step 1, we first define a set of periods to be analyzed by the wavelet based time-frequency analysis. For this sake, we use the linspace function of the Numpy package:

   ```
   periods = np.linspace(start=2*dt, stop=48, num=200)
   ```

   Above function generates an array of elements with values between the Nyquist period 2*dt and 48h in 200 equidistant steps.

3. We choose the time unit (here 'hours') and initialize the wavelet analyzer via

   ```
   wAn = WAnalyzer(periods, dt, time_unit_label='hours')
   ```

4. pyBOAT follows up on the pythonic idea of 'introspection', e.g. typing

   ```
   help(wAn)
   ```

   shows a comprehensive documentation of the WAnalyzer and its methods.

5. In order to show pyBOAT's results interactively we import the Python Matplotlib library via

   ```
   import matplotlib.pyplot as plt
   ```

6. Typing

   ```
   plt.ion()
   ```

   turns the interactive plotting mode on.

### 3.2.3 Detrending

Analogously to Section 3.1.5, we first detrend the raw time series using sinc filter smoothing as implemented within the pyBOAT package:

1. Define a sinc filter cut-off period in hours

   ```
   T_c = 48
   ```

2. The *sinc_smooth* function of the pyBOAT package requires two arguments, the raw signal that we aim to detrend as well as the cut-off period from point 1 of this section:

   ```
   trend = wAn.sinc_smooth(signal, T_c=T_c)
   ```

3. We subsequently obtain the detrended time series via subtracting the trend from the original signal

   ```
   detrended_signal = signal - trend
   ```

4. pyBOAT offers functions to plot the time series and corresponding trend:

   ```
   wAn.plot_signal(signal, label='Raw signal', color='red', alpha=0.5)
   wAn.plot_trend(trend, label='Trend')
   ```

   Note that any arguments (e.g. 'label', 'color' or 'alpha') that are accepted by the *plot* function of the *Matplotlib* package can be provided to above functions.

### 3.2.4 Computing the wavelet spectrum

1. We defined the parameters that are required for the wavelet analysis already in Section 3.2.2 step 3. We now perform the wavelet analysis on the detrended time series by typing

   ```
   modulus, transform = wAn.compute_spectrum(detrended_signal)
   ```

   This function also spawns a plot, showing the signal time-aligned with the wavelet power spectrum as obtained from the GUI (Figure 2A).

2. Here, "modulus" is a two-dimensional real valued array containing the Wavelet power spectrum while "transform" is a two-dimensional array representing the direct output of the complex convolutions with the Morlet mother wavelet. The number of rows equals the number of periods and the number of columns equals the length of the signal for both arrays.

3. Both objects *modulus* and *transform* are Numpy arrays and can be saved using e.g. the Numpy function *np.savetxt* or the Python *pickle* module.

### 3.2.5 Detect and evaluate the wavelet ridge

Analogously to Section 3.1.7, we compute and evaluate the maximal power ridge of the wavelet spectrum:

1. `ridge_results = wAn.get_maxRidge()`

   determines the maximum power ridge of the wavelet spectrum computed in Section 3.2.4 step 1 (*see* **Note 11**).

2. To plot the maximum ridge as a red line in the wavelet spectrogram type

   `wAn.draw_Ridge()`

3. The object 'ridge_results' defined in 3.2.5 step 1 contains the instantaneous period, phase, amplitude and maximum power values along the maximum power ridge of the wavelet spectrum. It is in a pandas DataFrame format such that the convenient I/O functions of the pandas software library can be used to save the results. For example,

   `ridge_results.to_csv("save_ridge.csv")`

   saves the results as a comma-separated values (csv) file, named "save_ridge.csv". Excel users might want to save their data into an Excel sheet using

   `ridge_results.to_excel("save_ridge.xlsx")`

4. These instantaneous properties can be plotted by typing

   `wAn.plot_readout(draw_coi=True)`

   which reproduces Figure 2B.

### 3.2.6 Import and analysis of experimental datasets

In this paragraph we study a previously published experimental data set. We show how such data can be imported within a Python scripting environment and subsequently analyze it.

**The data set:** Circadian rhythms are generated at the intracellular level through multiple interlocked regulatory feedback loops [14]. In mammals the negative feedback loop is composed of the *Period* (*Per1*, *Per2*, *Per3*), *Cryptochrome* (*Cry1*, *Cry2*) and *Bmal1* clock genes, see Figure 4A. This core loop is intertwined with multiple other loops such as the *Bmal1* and *Reverb* (*Reva* and *Revb*) negative feedback loop [15]. Recently, it has been shown that perturbations of the system given by jet-lag, light pulses, SCN slice preparations or culture medium exchange induce differential dynamical changes among different clock genes [16]–[20] and that these differential perturbation responses could be explained by the topology of the intracellular regulatory network [21]. Such differential responses translate into (at least transiently) different instantaneous amplitudes and periods among different clock genes and should thus be analyzed by a time-frequency analysis that can account for these complex and time-varying dynamical properties. In the next paragraphs we analyze *Bmal1-ELuc* and *Per1-luc* reporter expression within SCN slices of double-transgenic mice that express both reporters simultaneously as previously described in *Ono et al.* [19].

1. The data has been stored in a text-file "bioluminescence_raw.txt" containing three different columns of numerical values, i.e. the time instances of measurements as well as bioluminescence intensities of the *Bmal1-ELuc* and *Per1-luc* reporter constructs, respectively. The first row contains the data description, see Figure 4B.

2. There are multiple ways to load such data within a Python environment. One of the easiest is to use the *loadtxt* function of Numpy:

   ```
   import numpy as np
   ```

   ```
   np.loadtxt("./bioluminescence_raw.txt", skiprows=1)
   ```

   ```
   t, Bmal1, Per1 = Data.T
   ```

   Here, the file "bioluminescence_raw.txt" has to be in the same folder as the Python script. An alternative way is to use the convenient *read_csv* function of the Pandas library which is especially well suited for large data sets:

   ```
   import pandas as pd
   ```

   ```
   Data = pd.read_csv("./bioluminescence_raw.txt", sep="\t")
   ```

   ```
   t       = Data["Time"].values
   ```

   ```
   Bmal1   = Data["Bmal1"].values
   ```

   ```
   Per1    = Data["Per1"].values
   ```

   Figure 4C depicts the raw bioluminescence time series data as measured experimentally.

Note that the *Bmal1-ELuc* reporter has a much brighter overall light intensity but a smaller relative amplitude in comparison to the *Per1-luc* signal.

3. pyBOAT allows to detrend the raw time series signal and compute the wavelet spectrum in a single step which we will showcase in the next points.

4. Define the experimentally used sampling interval of the data, choose the corresponding time unit, set-up the periods of interest and initialize the wavelet analyzer via:

```
dt = 10./60    # Experimental sampling interval in hours
periods = np.linspace(2*dt, 48, 600)
wAn = WAnalyzer(periods, dt, time_unit_label='hours')
```

5. Define a cut-off period for sinc detrending, compute the wavelet spectra and determine the maximum power ridge for the *Bmal1-ELuc*

```
T_cutoff   = 96 # Define cutoff period in h
B1_modulus, B1_transform = wAn.compute_spectrum(Bmal1, T_c=T_cutoff)
B1_ridge = wAn.get_maxRidge()
```

and *Per1-luc* data

```
P1_modulus, P1_transform = wAn.compute_spectrum(Per1, T_c=T_cutoff)
P1_ridge = wAn.get_maxRidge()
```

compare with Section 3.2.3 and 3.2.4. Here, providing a parameter *T_c* within the *compute_spectrum* function of the wavelet analyzer allows to sinc-detrend and compute the spectrum within a single command. The detrended bioluminescence intensities of *Bmal1-ELuc* and *Per1-luc* reporters clearly show differential dynamics of oscillatory phases, i.e. an internal dynamical dissociation, see Figure 4D.

6. To further quantify this effect, we depict the instantaneous time-dependent periods from the *Bmal1-ELuc* and *Per1-luc* reporter as accessed via

```
Bmal1_period = B1_ridge["periods"]
Per1_period  = P1_ridge["periods"]
```

in Figure 4E. It can be seen that the *Bmal1-ELuc* oscillation period is slower than the *Per1-luc* period but then speeds up to values smaller than those of *Per1-luc*. Towards the end of the experiment, both periods approach each other again, indicating a rather transient dynamical dissociation followed by a subsequent resynchronization.

7. The differential period responses naturally translate into differential phase dynamics. We calculate the phase difference between the *Bmal1-ELuc* and *Per1-luc* oscillations,

```
from numpy import arctan2, sin, cos
Phasediff = arctan2( sin(P1_ridge["phase"] - B1_ridge["phase"]),
cos(P1_ridge["phase"] - B1_ridge["phase"]) )
```

using a distance metric that accounts for the cyclic nature of phase variables as previously described [22]. It can be seen that the initial large phase gap between *Per1-luc* and *Bmal1-ELuc* oscillations evolves towards an antiphasic relationship at around t=10d and ultimately saturates at a smaller phase difference of about 115°, see Figure 4F.

8. In conclusion, our wavelet-based time-frequency analysis helps to identify complex differential dynamical features in clock gene expression after SCN slice preparation and *in vitro* culturing.

## 4. Notes

1. The *Synthetic Signal Generator* allows to simulate oscillatory time series signals composed of the superposition of two non-stationary oscillations (so-called 'chirps') with different period behavior, an exponential decay as well as AR1 noise. Setting the AR1 parameter to zero corresponds to Gaussian white noise.

2. The wavelet analysis routine of pyBOAT expects equidistant time series sampling. Gaps in the recording can be interpolated, as described in the next point.

3. If there is missing data (gaps) in the time series pyBOAT offers a simple linear interpolation in between existing data points. See the GUI tooltip for the 'Set missing values entry' for the set of default characters encoding missing data (e.g. 'NaN') or define your own. Note that stretches of missing data at the beginning or end of a signal can only be interpolated to constant values.

4. The sinc detrending filter as implemented in pyBOAT, acts like a step function in the period domain, i.e. periods components of a signal that are below a certain threshold or cutoff-period are neither attenuated nor amplified while period components above the threshold can be related with the trend of the signal. Since the sinc filter has a non-zero roll-off in practical implementations for finite time series length, one has to carefully choose the threshold. Here, we have chosen a cut-off period of 48h since it is significantly above the expected oscillatory time scale of ~24h and thus detrending does not perturb the oscillatory properties of the signal, while it keeps the filter 'flexible' enough to reliably remove the non-linear baseline trend.

5. The choice of the mother wavelet function has a strong impact on the outcome of the analysis regarding the absolute power values. pyBOAT uses Morlet mother wavelets as a default which is one of the most widely used mother wavelet and known to fit sinusoidal-like signals well.

6. The main oscillatory component extracted via the wavelet ridge is strongly linked to the results obtained by the Hilbert transform, which is another commonly used non-stationary signal analysis approach. The Hilbert transform however is very noise vulnerable and generally requires a pre-smoothing of data obtained experimentally. In addition, the phase extracted via the Hilbert transform is different from results obtained by a wavelet analysis as it is waveform dependent [23].

7. In order to decide on a meaningful power threshold that divides the background noise from the signal components of interest, one needs a good null model for the background noise spectrum. In case of a white noise null model, the (period or frequency independent) threshold at a 95% confidence level is three [8], i.e. parts of the signal with a wavelet power larger than this can be assumed as statistically significant oscillations. However, the background spectrum of biological signals can significantly deviate from the white noise model due to correlations present. In case a reasonable null model is missing, one can estimate an empirical background spectrum from the experimental data itself. For the sake of consistency and reproducibility, it is important to keep and report the same threshold for the whole analysis, using a given experimental setup.

8. pyBOAT's underlying mathematical analysis is based on convolutions, which inherently display edge effects visualized by the cone of influence (COI). For very short signals it is possible that the entire ridge is inside the COI. As shown in detail in *Mönke et al.* [8], the phase, power and amplitude estimates are unreliable in these cases. However, period estimates show only very minor deviations. As a rule of thumb, the signal should have a length of at least three oscillations.

9. For repetitive analysis of similar data sets one can fix the default analysis parameters (e.g. the sampling interval or cut-off period) via the '*Settings*' menu entry of the main window.

10. PER2::LUC oscillations show a strong decline in amplitude after TTX application. We have selected a relatively low sinc filter cut-off value of 36h that is close to the expected dominant period of ~24h. Thus, we have chosen a compromise between potential mild perturbations of the main oscillatory component (*see also* **Note 4**) and the ability to reliably detrend the signal around the sudden jumps in oscillatory amplitude due to TTX treatment (compare Figure 3).

11. Analogously to the GUI of pyBOAT one can provide two parameters for the determination of the maximum power ridge, i.e.

    ```
    wAn.get_maxRidge(power_thresh=10, smoothing_wsize=20)
    ```

    While *power_thresh* gives a minimum wavelet spectrum power for which the ridge is determined, *smooting_wsize* provides the window size of the Savityky-Golay filter for ridge smoothing.

12. If it is possible to record known non-oscillatory signals within the same experimental setting (e.g. a nuclear fluorescent marker), pyBOAT can show the time-averaged wavelet power distribution for the whole ensemble (Batch Processing -> Fourier Spectra Distribution). The powers of this empirical background spectrum allow for a good estimate of a sensitive ridge power threshold to robustly detect oscillations. As shown in Mönke *et al.,* the minimal power required to statistically classify as 'oscillation' is three times the background power.

**Funding:** Christoph Schmal's research was supported by the Deutsche Forschungsgemeinschaft (DFG) through grant SCHM3362/2-1. Gregor Mönke's research was supported by the EMBL Interdisciplinary Postdoc Programme (EIPOD) under Marie Slodowska-Curie Actions COFUND grant number 664726. Adrián E. Granada's research was supported by the German Federal Ministry for Education and Research (BMBF) through the Junior Network in Systems Medicine, under the auspices of the e:Med Programme (grant 01ZX1917C). The funders had no role in study design, data collection and analyses, decision to publish or preparation of the manuscript.

**Acknowledgements:** We gratefully acknowledge Daisuke Ono, Sato Honma, Ken-ichi Honma, John Abel and Erik Herzog for sharing experimental data as well as Carolin Ector for comments on our manuscript.

**Conflicts of interest:** The authors declare no conflict of interest.

# Figures

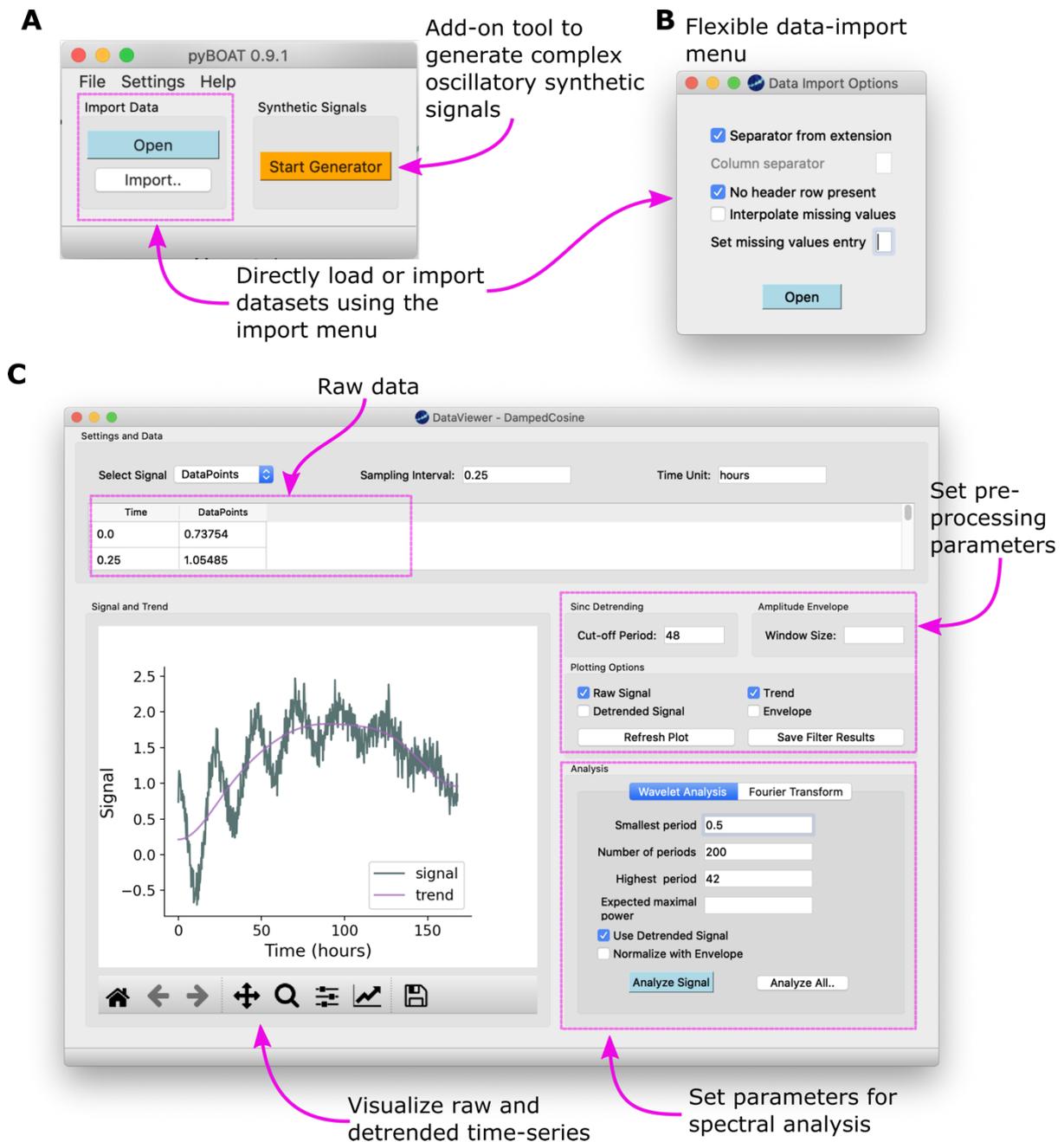

**Figure 1. Data import and parameter setup for the analytic wavelet transform.** A) Main window of pyBOAT. B) *Data Import Options* window. C) The *DataViewer* shows the imported data, plots the time series and trend of interest and allows to define parameters for sinc filter detrending and the subsequent wavelet analysis.

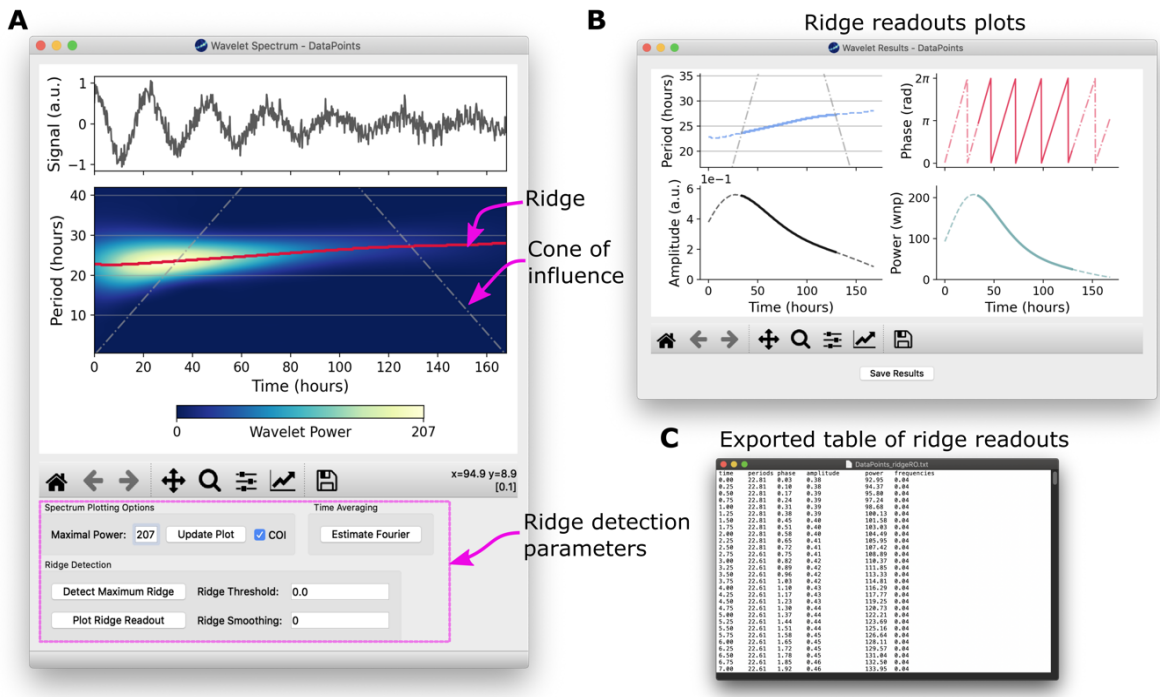

**Figure 2. Wavelet analysis and ridge readout.** A) *Wavelet spectrum* window. Upper window depicts the analysed signal, i.e. in our case the sinc-detrended time series. Bottom window depicts the wavelet spectrogram together with the detected maximum power ridge (*bold red line*) and the cones of influence (*grey dashed-dotted lines*) B) *Wavelet Results* window, obtained by clicking on the *Plot Ridge Readout* button in panel A. Depicted are the time-dependent (instantaneous) period (*upper left*), phase (*upper right*), amplitude (*bottom left*) and maximum power values (*bottom right*) as evaluated from the maximum power ridge in panel A. C) Comma separated value (csv) readout, obtained by clicking on the *Save Results* button in panel B.

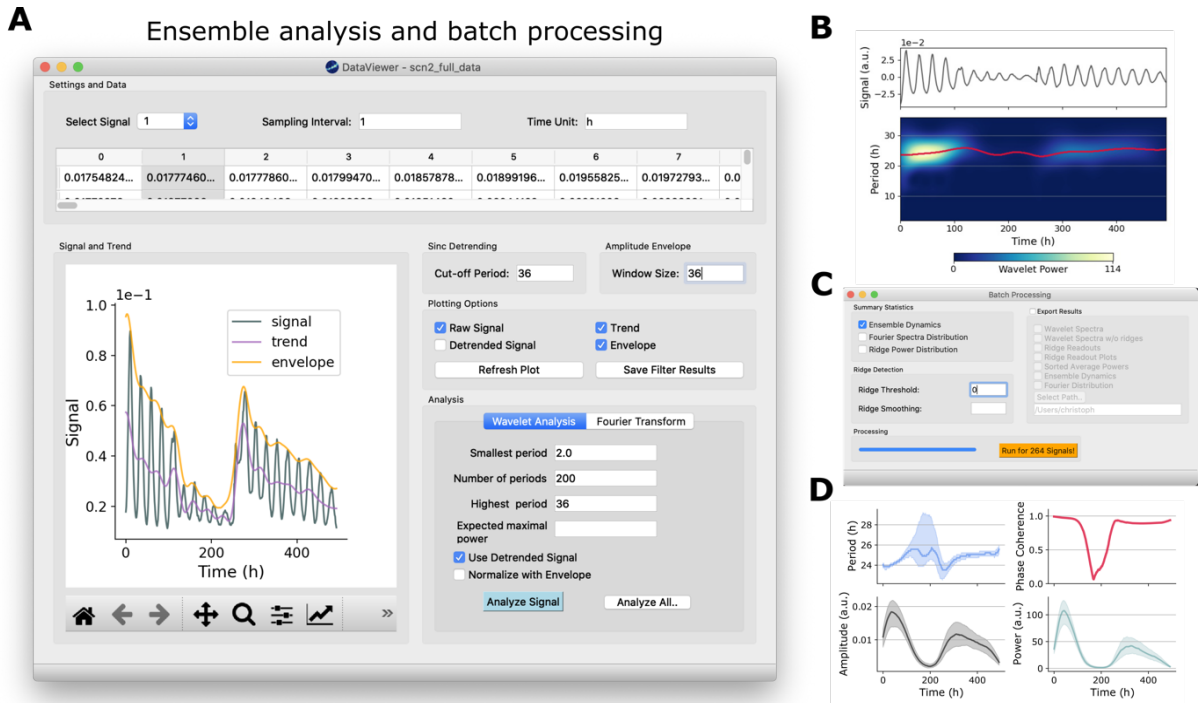

**Figure 3. Analysis of large data sets.** A) The *DataViewer* window of pyBOAT. A large data set has been imported that contains 264 bioluminescence time series signals obtained by individual tracking of SCN neurons within a cultured SCN slice at a sampling rate of 1h. B) Wavelet analysis of a bioluminescence recording from an individual SCN neuron stored in column one as shown in panel A. C) The *Batch Processing* window allows to specify options for the analysis of large ensemble data sets. D) Summary statistics of the corresponding ensemble analysis of all 264 recordings. Bold lines denote averages while shaded areas denote standard deviations of oscillation properties evaluated along the maximum power ridges of all signals.

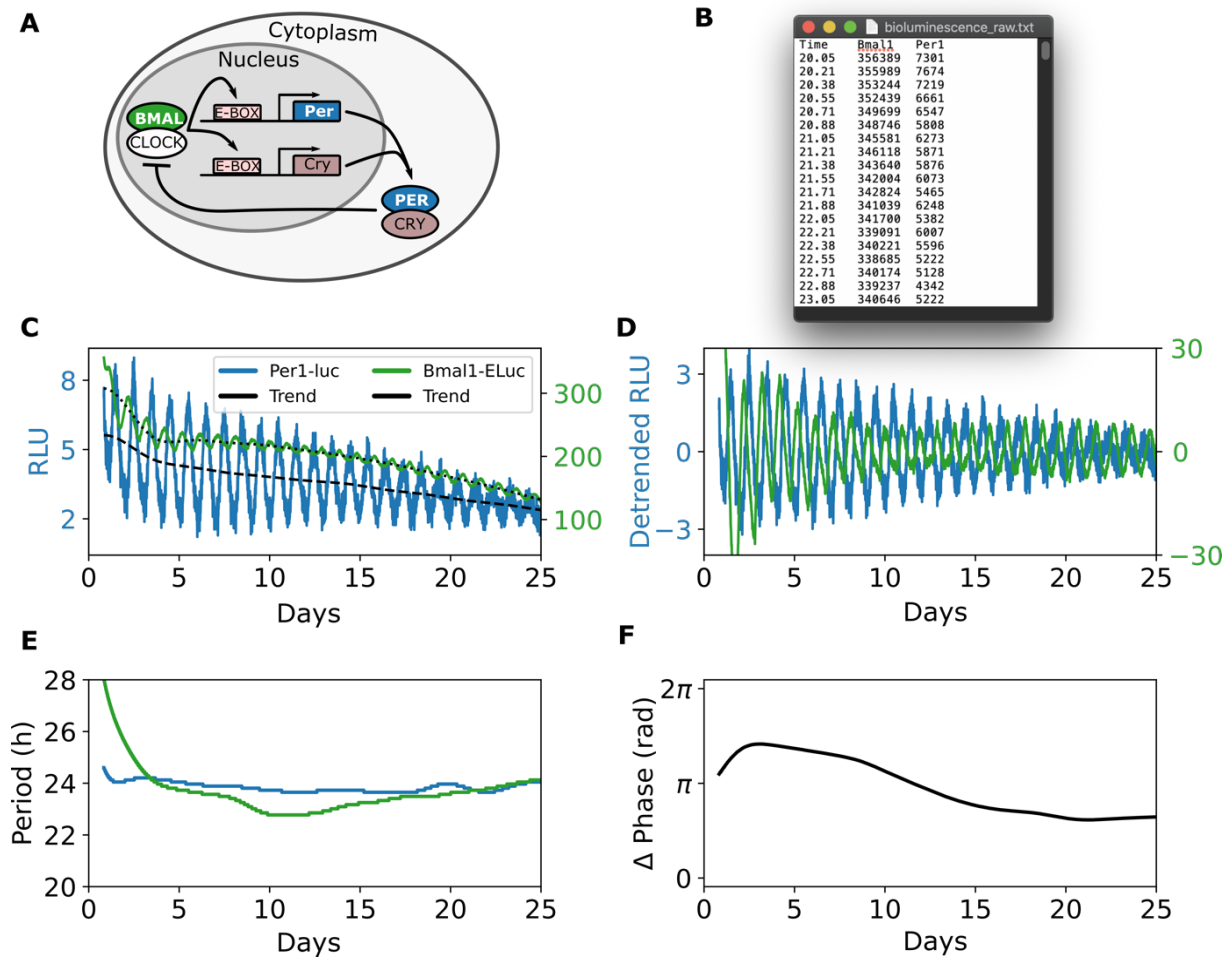

**Figure 4. Detecting differential phase responses of *Bmal1-ELuc* and *Per1-Luc* clock gene reporter expression.** A) Sketch of the mammalian circadian core clock regulatory network. B) Illustration of the data set. C) Raw non-detrended *Bmal1-ELuc* (*green*) and *Per1-luc* (*blue*) bioluminescence given in relative light units (RLU; 1RLU = 1000counts per 15 seconds), measured via a photomultiplier tube (PMT) as previously described [19]. Nonlinear baseline expression trends (magnitudes), determined by a sinc-filter, are depicted by a black dotted or dashed line in case of *Bmal1-ELuc* or *Per1-luc* signals, respectively. D) Corresponding detrended bioluminescence signals, calculated by subtracting the trend from the original raw time series data from panel (C). E) Time-dependent instantaneous periods, evaluated from the maximum power ridge of the corresponding wavelet spectrograms (data not shown). F) Difference of the time-dependent instantaneous *Bmal1-ELuc* and *Per1-luc* oscillation phases.